\documentclass[reprint,aps,amsmath,amssymb,superscriptaddress,prd,nofootinbib]{revtex4-1}
\usepackage{float}

\def\mH{{\mathcal H}}

\def\L{{\cal L}}

\newcommand{\be}[1]{\begin{equation}\label{#1}}
\newcommand{\ee}{\end{equation}}

\usepackage{mathtools}

\begin{document}
\title{Bondi-Metzner-Sachs invariance and electric-magnetic duality}
\author{Claudio Bunster}

\affiliation{Centro de Estudios Cient\'{\i}ficos (CECs), Casilla 1469, Valdivia, Chile}
\author{Andr\'es Gomberoff} 
\affiliation{Facultad de Ingenier\'ia y Ciencias, Universidad Adolfo Ib\'a\~nez, Avda.~Diagonal las Torres 2640, Pe\~nalol\'en, Santiago, Chile}
\affiliation{Centro de Estudios Cient\'{\i}ficos (CECs), Casilla 1469, Valdivia, Chile}
\author{Alfredo P\'erez}
\affiliation{Centro de Estudios Cient\'{\i}ficos (CECs), Casilla 1469, Valdivia, Chile}

\begin{abstract}

We exhibit a Hamiltonian formulation,  both for electromagnetism and gravitation, in which it is not required that the Bondi ``news" vanish, but only that the incoming news be equal to the outgoing ones. This requirement is implemented by defining the fields on a two--sheeted hyperbolic surface, which we term ``the hourglass". It is a spacelike deformation of the complete lightcone. On it one approaches asymptotically (null) past and future infinity while remaining at a fixed (hyperbolic) time, by going to large spatial distances on its two sheets. The Hamiltonian formulation and - in particular - a conserved angular momentum, can only be constructed if one brings in both, the electric and magnetic BMS charges, together with their canonically conjugate ``memories". This reveals a close interplay between the BMS and electric-magnetic duality symmetries.
\end{abstract}

\maketitle

\section{Introduction}
\setcounter{equation}{0}

The connection between radiation and the existence of an asymptotic infinite dimensional symmetry algebra, discovered by Bondi, Metzner and Sachs (BMS)\cite{Bondi:1962px,Sachs:1962zza,Sachs:1962wk} has been a fascinating subject ever since it emerged. In recent years attention on it was revived due first to the work of Barnich et al.\cite{Barnich:2001jy,Barnich:2009se, Barnich:2011mi}, and later through that of Strominger et al.\cite{Strominger:2013jfa,Strominger:2017zoo}. Barnich employed a classical Hamiltonian defined on future lightcones, and a construction termed by him a ``covariantized Regge--Teitelboim method" which needed a spacelike deformation of the lightcone in order to define a Poisson bracket. Strominger, on the other hand, was basically quantum mechanically inclined and also worked on lightcones. Among his many contributions, he took - already at the classical level - an important new step by bringing together future and past-like cones, joined through a spatial inversion that he termed the ``antipodal map". He also brought new light, this time at the quantum level, onto the so called ``angular momentum problem", which is the name that was given in the past to the existence of many angular momenta connected with each other by BMS transformations. His response was to feel at ease with the ``problem" by stating that the transformation that mapped one angular momentum to the other, connected different vacua.

The work reported herein takes element from both of the above developments. First it introduces a spatial spacelike deformation, but this time of the {\it complete} lightcone, past and future, that we have termed ``the hourglass" because of its shape. This two-sheeted surface, which automatically incorporates the antipodal map is, however, not brough in as an auxiliary device to define Poisson brackets, but rather as a fundamental ingredient: it is the surface on which the fields are defined {\it instead} of the lightcone. 

Since it is spacelike, the hourglass has the advantage of enabling one to use the standard, battle tested, Regge-Teitelboim procedure\cite{Regge:1974zd} to define the Hamiltonian. On it one approaches asymptotically (null) past and future infinity while remaining at a fixed (hyperbolic) time, by going to large spatial distances on its two sheets.

If one constructs the Hamiltonian by ``improving" the generators of different motions so they have well defined functional derivatives, one finds that this can only be done if one brings in both, the electric and magnetic BMS charges together with their canonically conjugate ``memories". This reveals a close interplay between the BMS and electric-magnetic duality symmetries. This interplay becomes specially poignant in connection with angular momentum, which can only be defined so that it is conserved, with the help of electric-magnetic duality invariance. 

The construction of the Hamiltonian in the presence of radiation also confirms, in a blatant manner, the crucially different role of ``improper gauge transformations", whose generators involve surface integrals from that of the proper ones whose generators do not. The former are to be regarded as changing the physical state, and are not trivial symmetries, while the latter are just due to redundant counting, and can be factored out by taking a quotient or by fixing the gauge.

In the present case one finds that, already for electromagnetism, what one thought was an ``internal symmetry" is inextricably intertwined with spacetime displacements due to the ``memory" carried by ``the news". And this implies that for the angular momentum there is no ``problem" if one simply accepts what the theory is expressing each and every time it is able to: improper gauge transformations change the physical state. And this is already seen at the classical level.

The plan of the paper is the following. Section II introduces the hourglass and discusses its properties, then section III develops the formalism for the electromagnetic field. Finally section IV is devoted to gravitation.

In the case of gravitation, an electric-magnetic duality invariant description of the linearized theory on the hourglass has not yet been developed; but one can guess by analogy some of its elements. The proposals of that section concerning magnetic BMS charges are, therefore, of a speculative nature.

The results presented in this paper were obtained while improving a manuscript that had been elaborated for a book in preparation, in honor of Tulio Regge\cite{book}, and are being incorporated in its updated version. This permits us to use that extensive report as an overall reference and review, and concentrate herein on the conceptual issues avoiding technicalities as much as possible. 

\section{The hyperbolic hourglass}

The hyperbolic hourglass consists of an outgoing hyperboloid with center $x^\mu_{(0)}=(-\tau_0,0)$ joined 
 to an incoming one  with center $x^\mu_{(0)}=(\tau_0,0)$.
 It obeys,
\begin{equation}
(x^{\mu}-x_{(0)}^{\mu})(x_{\mu}-x_{(0)\mu})=-\tau_{0}^{2}\, ,\label{hyp}
\end{equation}
and it is defined parametrically from Minkowskian coordinates $(x^0,\vec x)$ through
\begin{eqnarray}
x^{0} & = & t+ r\sqrt{1 + \frac{\tau_0^2}{ r^2}} + \tau_0  \ \ \ r\leq 0 \label{emb1}\\
x^{0} & = & t+ r\sqrt{1 + \frac{\tau_0^2}{ r^2}} - \tau_0    \ \ \ r\geq 0, \label{emb2}
\end{eqnarray}
\be{x}
\vec{x} = r \hat{r}\ , 
\  \ -\infty<r<+\infty,
\ee
where the unit vector $\hat r$ is given by
\be{unitr}
\hat r = (\cos\vartheta\cos\varphi,\cos\vartheta \sin\varphi, \sin\vartheta),
\ee
where
\begin{eqnarray*}
 0\leq &\vartheta&\leq\pi, \\ 0\leq &\varphi&<2\pi.
\end{eqnarray*}
The radius $\tau_0$ is taken to be positive.  The embedding defined by the above equations is continuously differentiable. The tangent vectors are continuous at $r=0$ and the surface has a well defined global orientation. 

The hyperbolic hourglass may be regarded as a spacelike deformation of the full (pass and future) lightcone, with an orientation inherited from the propagation of a light front that comes in, goes through itself, and then comes out. Since this wave propagation process is physically smooth, fields defined on the global coordinate system just described should be smooth.

 The parametric equations   \eqref{emb1}-\eqref{x} automatically incorporate the antipodal map \cite{Strominger:2013jfa} \cite{Strominger:2017zoo},  which amounts to rewriting them by using a positive $r$ for both sheets of the hyperboloid and inverting the orientation of the two-spehere at a given $r$.  That is, keeping   \eqref{emb1}-\eqref{x} for $r\geq 0$ and setting,
	$r' = -r$, $\hat r' = -\hat r$, for $r\leq 0$.
	
	If one considers an incoming wave which is not spherically symmetric, then the spacetime point at which the wavefront goes through itself will be different for different $\hat r$'s. But in the present paper we are only interested in the analysis of the asymptotic region and therefore the details of what happens inside are irrelevant. The key aspects are the asymptotic hyperbolic shape and its orientation inherited from that of an incoming wave that goes through itself and becomes outgoing.

	Figure 1 shows the embedding in  Minkowski space of a single hyperbolic hourglass, figure 2 exhibits the slicing of Minkowski space by a one parameter family of hyperbolic hourglasses and figure 3 shows a sequence of Penrose diagrams with hyperbolic slicings of different radius $\tau_0$.
	
	The hourglass foliation consists of hyperboloids of fixed radius and varying center. In contradistinction, hyperbolic foliations used previously by several authors have had fixed center and varying radius\footnote{See, for example,
\cite{Ashtekar:1978zz}, \cite{0264-9381-9-4-019},  \cite{Campiglia:2015qka},
and also  \cite{Troessaert:2017jcm} and references
therein. In some of these discussions timelike hyperboloids are employed
(in which \nopagebreak case $-\tau_{0}^{2}$ in \eqref{hyp} is replaced by $\lambda_{0}^{2}$).} .
\begin{figure}[H]\label{hourglassfig}
\begin{center}
\includegraphics[width=8cm]{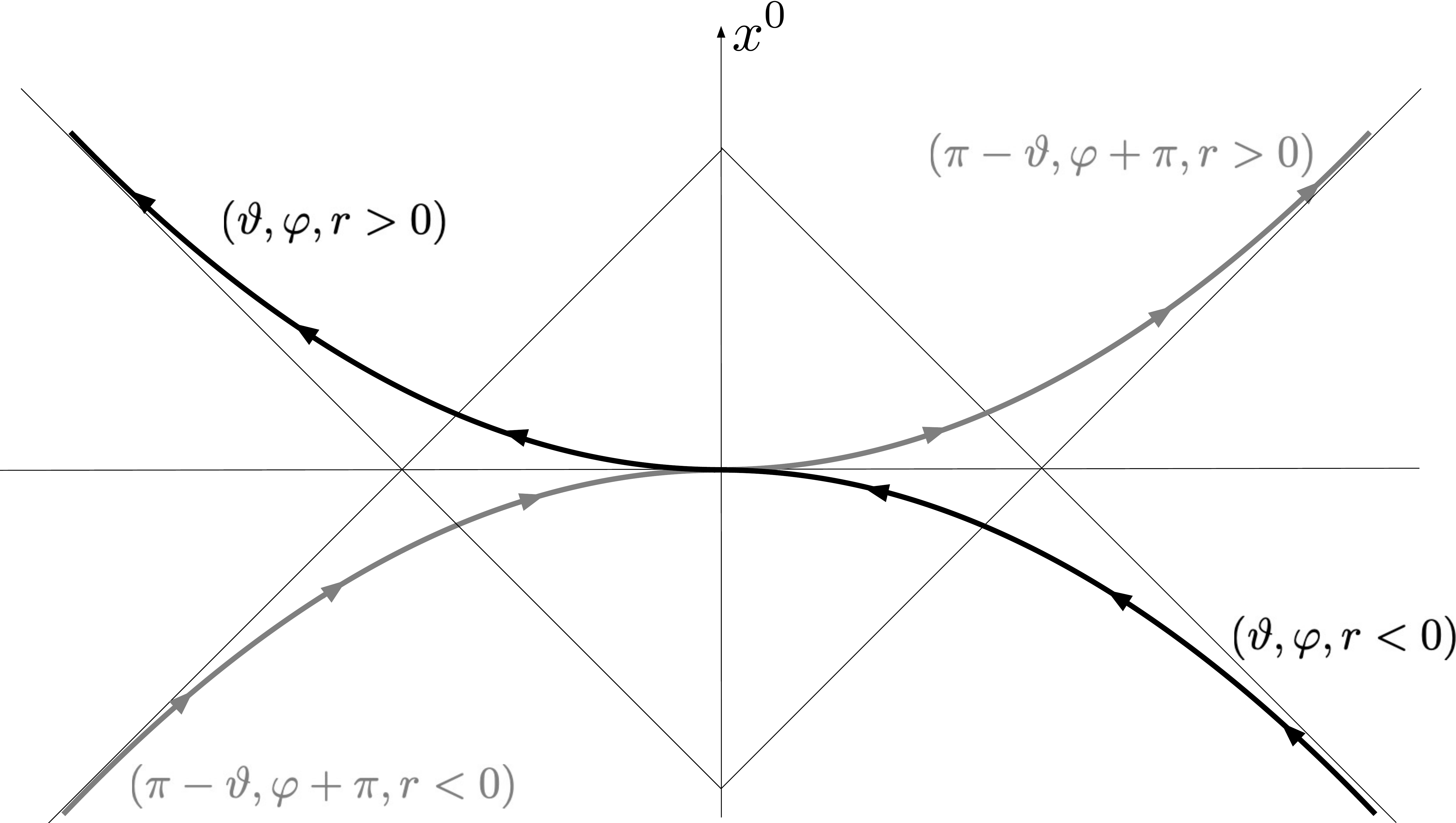}
\caption{{\it The hyperbolic hourglass}.
The figure shows a two dimensional cut of an incoming hyperboloid and an outgoing one which are joint smoothly at at $r=0$. The arrows show the direction of increasing $r$, which coincide asymptotically with the direction of propagation of a wave that comes in, goes through itself, and then comes out.
 If the incoming wave is not spherically symmetric, the spacetime point at which the wavefront goes through itself will be different for different $(\vartheta,\varphi)$. For the analysis of the asymptotic region the details of what happens inside are irrelevant. The key aspects are the asymptotic hyperbolic shape and its orientation inherited from that of an incoming wave that goes through itself and becomes outgoing.}
\end{center}
\end{figure}
\begin{figure}[H]\label{twopa}
\begin{center}
\includegraphics[width=7cm]{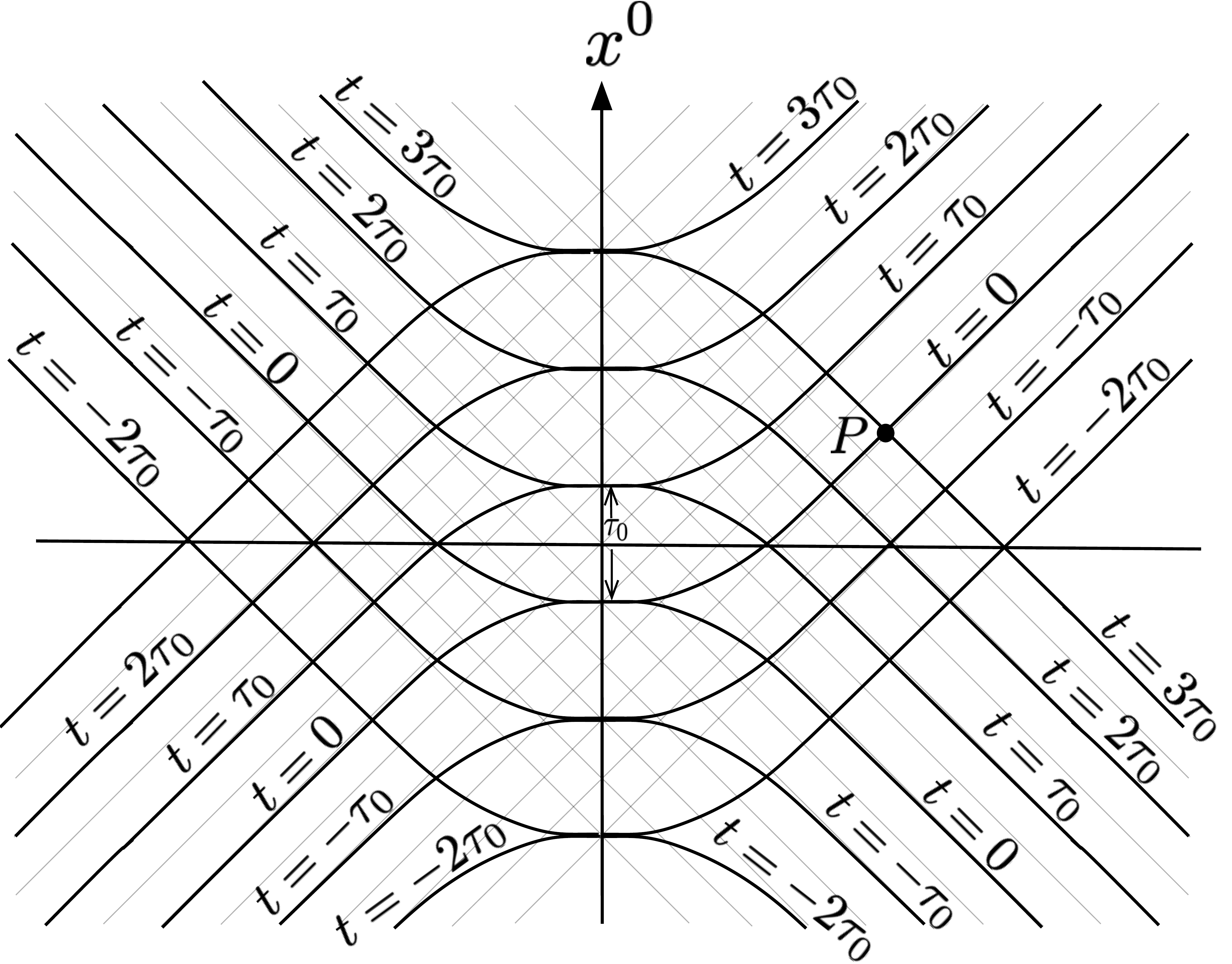}
\caption{{\it Slicing by hyperbolic hourglasses}. A given spacetime point is labeled by two set of
 coordinates. In the case of the point $P$ shown in the figure, these are $(t=0, r, \vartheta,\varphi)$ 
 and $(t=3\tau_0, -r, \pi-\vartheta,\varphi+\pi)$.}
\end{center}
\end{figure}
\begin{figure}[H]
\begin{center}
\includegraphics[width=8cm,height=8cm]{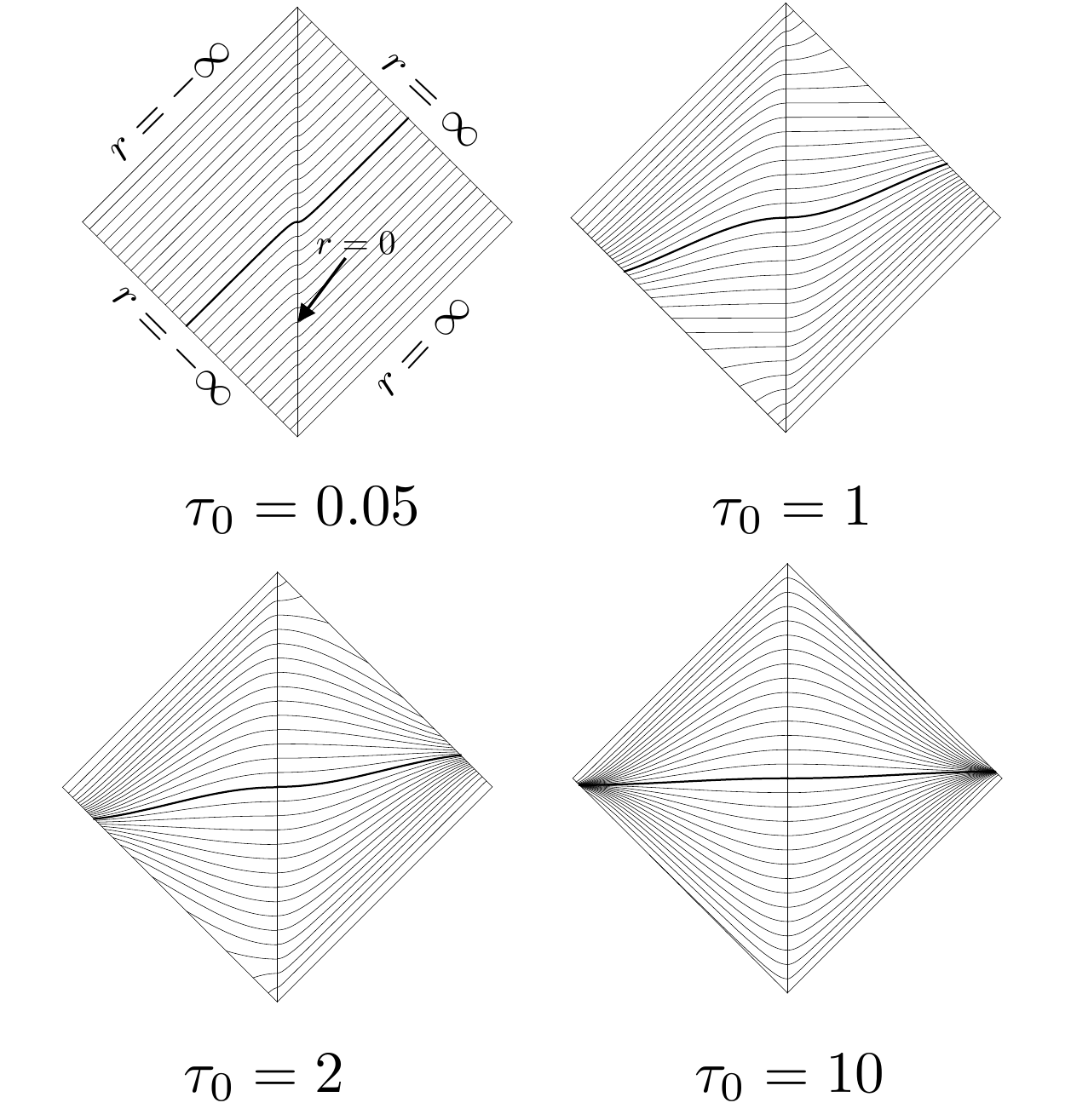}\label{mat}
\caption{{\it Limits $\tau_{0}\rightarrow0$ and $\tau_{0}\rightarrow \infty$ for Minkowski space}.
The succession of conformal diagrams shows from left to right how the surfaces of the  
hourglass foliation are
deformed from nearly light cones to nearly planes as $\tau_{0}$ increases from a very small 
value to a very large one. To better illustrate the effect, different members of  the foliation are
 shown in the different figures of the sequence; but, to keep 
track of the deformation, the surface at $t=0$ (shown with a heavy line) in all cases. 
The Penrose diagram has been doubled to admit negative values of $r$ in the left 
triangular area. This doubling shows how the curves of constant $t,\vartheta,\varphi$ 
are smooth spacelike curves that connect asymptotically past and future null 
infinities.
 The scale of the lenght $\tau_0$ is irrelevant for the effect described in the figure, which only
  depends on the ratios between the different $\tau_0$'s shown.}
\end{center}
\end{figure}

%%%%%%%%%%%%%%%%%%%%%%%%%%%%%%%%%%%%%%%%%%%%%%%%%%%%%%%%%%%%%%%%%%%
%%%%%%%%%%%%%%%%%%%%%%%%%%%%%%%%%%%%%%%%%%%%%%%%%%%%%%%%%%%%%%%%%%%
\section{Electromagnetic field in Minkowski space}

\label{sec:EMMink}
%%%%%%%%%%%%%%%%%%%%%%%%%%%%%%%%%%%%%%%%%%%%%%%%%%%%%%%%%%%%%%%%%%%
%%%%%%%%%%%%%%%%%%%%%%%%%%%%%%%%%%%%%%%%%%%%%%%%%%%%%%%%%%%%%%%%%%%

We will analyze in this section the case of the electromagnetic field
on a fixed Minkowskian background. Practically all the features that
will be encountered in the gravitational case already appear in this
technically simpler context. 

The main difference, which does not hinder the analogy, is that, since the background is fixed, its Poincar\'e symmetry appears as a global
symmetry rather than an asymptotic gauge symmetry. There are no constraints
associated with the surface deformation $\xi$, which are not varied
in the action principle. The Hamiltonian is
\begin{equation}
H_{0}\left[\xi\right]=\int d^{3}x\left(\xi^{\perp}\mH_{\perp}^{(elm)}+\xi^{i}\mH_{i}^{(elm)}\right)\,,\label{H0}
\end{equation}
where the $\mH_{\mu}$ in \eqref{H0} are replaced
by the energy and momentum densities of the electromagnetic field,
\begin{eqnarray}
\mH_{\perp}^{(elm)} & = & \frac{1}{2}\left(g^{-\frac{1}{2}}\pi_{i}\pi^{i}+\frac{1}{2}g^{\frac{1}{2}}F^{ij}F_{ij}\right)\,,\label{hpe}\\
\mH_{i}^{(elm)} & = & F_{ij}\pi^{j}\,,\label{hie}
\end{eqnarray}
and $\xi^\perp$ and $\xi^i$ may be traken to be the normal and tangential components of any of the Poncar\'e Killing vectors. 

The only gauge symmetry present in the problem is the electromagnetic
one, whose generator is 
\begin{equation}
{\cal G}=-\pi_{,i}^{i}\approx0\,.\label{gg}
\end{equation}
Here $A_{i}$ is the vector potential, $\pi^{i}$ its conjugate momentum,
$g_{ij}$ is the metric on the hourglass, and $g$ denotes its
determinant.

If instead of having a fixed background we were considering dynamically
coupled electromagnetic and gravitational fields, then expressions
\eqref{hpe}, \eqref{hie} would be added to their gravitational counterparts 
discussed in section \ref{sec:Gravfield}, 
and the sum would be constrained to vanish. The asymptotic analysis given
below would still hold because at large distances the spacetime would be flat. Then the asymptotic
symmetry transformations of the coupled Einstein-Maxwell system would be those discussed
here (internal electromagnetic, and Poincar\'e transformations) and the additional
gravitational supertranslations.

We will now discuss the Poincar\'e and proper and improper gauge transformations
for the electromagnetic field on the hourglass slicing. In this case
the time equal constant surface is left invariant under the Lorentz
group, whereas it is mapped onto a different hourglass by spacetime
translations. Thus if one compares the situation with $t=$ constant
planes, one sees that the roles of spatial translations and boosts
are interchanged.

\subsection{Asymptotic boundary conditions}
\subsubsection{Power expansion near $r=\pm \infty$}

Starting from the Coulomb field written in hyperbolic coordinates,
one is led to the boundary conditions, 
\begin{eqnarray}
A_{a} & = & a_{a}^{(0)}+\mathcal{O}(r^{-1})\,,\label{a1}\\
A_{r} & = & a_{r}^{(2)}r^{-2}+\mathcal{O}(r^{-3})\ ,\label{a2}\\
\pi^{a} & = & \pi_{(2)}^{a}r^{-2}+\mathcal{O}(r^{-3})\,,\label{a3}\\
\pi^{r} & = & \pi^r_{(0)}+\mathcal{O}(r^{-1})\, , \label{a4} \ \ \ \ \ \ \ \ \label{a4}\\
\lambda &=& \lambda_{(0)}+\mathcal{O}(r^{-1}) ,\label{gauges}
\end{eqnarray}
 Here $\lambda$ is the Lagrange multiplier that accompanies the gauge generator \eqref{gg}. In addition to the power law decays \eqref{a1}--\eqref{gauges} it is necessary to introduce parity conditions. This is achieved by splitting some of the variables in longitudinal and transverse parts as follows
\begin{eqnarray}
a^{(0)}_{a}=\nabla_{a}F+\star\nabla_{a}\bar{G}\, ,\label{eq:divcurl1} \\
h_{a}=\gamma^{\frac{1}{2}}\left(\nabla_{a}N+\star\nabla_{a}\bar{N}\right)\, .\label{eq:divcurl2}
\end{eqnarray}
Here, 
\begin{equation}
\star\nabla_{a}=\gamma^{\frac{1}{2}}\epsilon_{ab}\nabla^{b}\,\label{nablabar},
\end{equation}
where $\gamma$ is the determinant of the metric $\gamma_{ab}$ on the unit two-sphere. The ``news'' vector $h^a$ in \eqref{eq:divcurl1}, which will play a central role in what follows, is defined by
\begin{equation}
h^{a}=\frac{1}{\tau_{0}^2}\left(\pi^{a}_{(2)}+\gamma^{\frac{1}{2}}\gamma^{ab}f_{b r}^{(2)}\right)\, ,\label{ha}
\end{equation}
for $r\rightarrow\pm\infty$, where the
 $f^{(2)}_{br}$ is the leading order coefficient of $F_{br}$.
In Minkowski coordinates the news correspond to an electromagnetic
field that decays as $r^{-1}$, that is to a wave emerging from a confined source $(r\rightarrow+\infty)$, or converging towards an absorber$(r\rightarrow-\infty)$. For an accelerating electric charge $e$ one has from the Lienard-Wiechert field,
$$
\gamma^{\frac{1}{2}}f^{ar}_{(2)}\vec{\partial}_{a}=\pi_{\left(2\right)}^{a}\vec{\partial}_{a}=-2e\gamma^{\frac{1}{2}}\hat{r}\times\left(\hat{r}\times\vec{a}\right),
$$
where  $\vec{a}$ is the acceleration
 in rest frame of the emitter (outgoing wave) or absorber (incoming wave).
See for example, \cite{Rohrlich,Teitelboim:1970mw}. 

\subsubsection{Parity conditions}

The parity conditions will be the following,
\be{parity}
\left. (F,\lambda_{(0)}, N,\bar N)\right|_{r=+\infty} =
\left. (F,\lambda_{(0)}, N,\bar N)\right|_{r=-\infty}  ,
\ee
 for each $(\vartheta,\varphi)$.
 
 Parity conditions play a fundamental role in the Regge-Teitelboim discussion of Poincar\'e invariance
on asymptotic planes. We see that when dealing with Bondi, Metzner, Sachs invariance on hyperboloids, in the presence of news,
they again come in\footnote{The BMS symmetry has been tamed to fit a foliation by surfaces that are
asymptotically planes  \cite{Henneaux:2018cst,Henneaux:2018gfi,Henneaux:2018hdj,Henneaux:2019yax}.
This has required dexterity, since the symmetry is intimately related to radiation and
its natural habitat is an asymptotically null surface, rather than a plane.}. It is shown in \cite{book} that the boundary conditions (power expansion and parity requirement) are preserved under Poincar\'e and improper gauge transformations.

The physical motivation for the parity conditions is very simple. They state that for a closed system (the free electromagnetic field in this case) everything that comes
in must come out. That is, one allows for non--vanishing incoming and outgoing fluxes of energy,
momentum, and other (BMS) charges; but requires that the net flux should be equal to zero.

This requirement, which physically is a condition connecting the remote past with the remote future,
can be formulated as a fixed time statement, because the spacelike hyperbolic hourglass is
asymptotically tangent to the past and future lightcones. This is the reason for bringing it in to begin with.

\subsection{The hyperbolic hourglass as an unconventional Cauchy surface}

When regarded as an initial value surface, the hourglass has the unconventional 
feature, that a spacetime point, which is not at infinity, lying, say, on the outgoing half of
 one hourglass at a given time, also lies on the incoming half of another hourglass at a later time. 
 This implies that one cannot give freely initial value data on the complete hourglass but only on half of it,
 the outgoing half for example. However, the double ocurrence of points does not happen at infinity, so if one 
 gives data on the outgoing half one should specify additionally the incoming radiation, that is one should give
 the news at $r=-\infty$. But this is precisely what the parity condition does, stating that the 
 incoming news are equal to the outgoing ones. Thus it is sufficient to specify just the data on the outgoing
  half of the hourglass (or, viceversa, on the incoming one) if the parity condition is imposed.

Therefore one must bring in the complete hourglass in order to deal in Hamiltonian terms with the 
interrelationship
 between past and future, but one only gives initial value data on one half of it, together with asymptotic 
 information on the other half. In this sense the hourglass plays the role of a Cauchy surface.

\subsection{Fiber memory}
\label{sec:Poinc}

For a pure time translation the equation of motion for the leading order term  of $A_a$ is,
 \be{dotAa}
 \dot a_a^{0}=\gamma^{-\frac{1}{2}}h_a,
\ee
 for $r\rightarrow\pm\infty$. Its longitudinal component is
\be{Fdott}
\dot F =N.
\ee

Equation \eqref{Fdott} has a highly non trivial content.
It shows that, even when the generator of improper gauge transformations does not act, i.e., when  $\lambda_{(0)}=0$, and one is only moving in the time $t$, there is still a displacement,
\be{Ndt}
\delta F=N\delta t 
\ee
along the U(1) fiber at each $(\vartheta,\varphi)$, of amount $N\delta t$, when a time $\delta t$ elapses. That is: (i) If there are no news (and one does not change the gauge frame) $F$ is conserved, (ii) If there are news during a time interval the value of $F$ changes from $F_{\mbox{\tiny before}}$ to $F_{\mbox{\tiny after}}$ according to the integral of  \eqref{Fdott} over the time interval. That is, $F$ ``remembers'' the news, and for that reason is called the ``fiber memory''. Another kind of memory, ``charge memory'' will be encountered below in section \ref{ssrates}.

It is important to realize that \eqref{Ndt} is not just a ``redefinition of $\lambda_{(0)}$ by the amount $N$". This is because $\lambda_{(0)}$ is not in the phase space, and can be held fixed in the variation of the Hamiltonian, whereas $N$ is a dynamical variable, which obeys a (gauge invariant) equation of motion and hence cannot be held fixed.

\subsection{BMS charges}

\subsubsection{Electric BMS charge}

Taking into account the parity condition on $\lambda_{(0)}$ one finds that the surface integral that must be added to the electromagnetic gauge generator to include improper transformations is given by
\be{gaugeimp}
\oint \lambda_{(0)} Q
\ee 
where the gauge charge $Q$ is given by
\begin{equation}
Q\left(\vartheta,\varphi\right)=\left.\pi_{(0)}^{r}\right|_{+\infty}-\left.\pi_{(0)}^{r}\right|_{-\infty} \equiv Q_+ + Q_-.\label{eq:Qgauge}
\end{equation}
It is important to interpret this expression appropriately. The hourglass is a construct that enables one to keep track, within the Hamiltonian formalism, of the incoming and outgoing radiation in an economic manner, that is  without introducing separate overlapping incoming and outgoing hyperbolic patches. This brings in a redundancy: one way or another space ocurrs twice. We just saw one instance of this above in connection with the initial value data. The redundancy strikes again in expression \eqref{eq:Qgauge} for the charge. If one considers the Coulomb field of a particle of charge $e$ at rest at $x^i=0$ one finds,
$$
Q_+\left(\vartheta,\varphi\right) =\frac{e}{4\pi}\sin\vartheta
$$
and
$$
Q_- \left(\vartheta,\varphi\right)=\frac{e}{4\pi}\sin\vartheta
$$
and hence
\be{Qcoul}
Q\left(\vartheta,\varphi\right)= 2\frac{e}{4\pi}\sin\vartheta
\ee
The factor two arises because one is counting twice: $Q_+$ is the charge as seen in the
 outgoing description of space, while $Q_-$ is the {\it same} charge as seen from its incoming replica.
  This point will reappear below in connection with radiation rates.

\subsubsection{Magnetic BMS charge}

There is a magnetic analog of \eqref{eq:Qgauge} given by 
\be{barQ}
\bar Q=\left.\epsilon^{ab}\nabla_a a^{(0)}_{b} \right|^{+\infty}_{-\infty}=\left.\gamma^{1/2}\nabla^2\bar G\right|^{+\infty}_{-\infty} ,
\ee
which is conserved as a consequence of \eqref{dotAa} and the parity condition 
for $\bar{N}$, 
 \be{dQ0}
 \dot{\bar Q}=0,
 \ee

 In the electric representation this conservation law  appears as an ``accidental'', because it does
  not follow from a symmetry of the action. The formalism becomes complete if one introduces
   a second potential, so that the electric and magnetic charges are treated on the same footing.
    This completion of the formalism may be regarded as a matter of elegance and economy, but not
     of necessity, for questions that can be asked within the electric representation. But, as we will see
     further below, it becomes essential when one discusses Lorentz transformations. Therefore we recall it right
       away.

%%%%%%%%%%%%%%%%%%%%%%%%%%%%%%%%%%%
\subsection{Asymptotic two potential formulation}

%%%%%%%%%%%%%%%%%%%%%%%%%%%%%%%%%%%%%%%%%%%%%%%%%%%%%%%%%%%%%%%%%%
\label{mm}

One brings in a new, 
``magnetic" vector potential $\bar{A}$. For the present purposes it is sufficient to do so only asymptotically.  The potential $\bar{A}$ satisfies, 
\be{abardef}
\pi^{r}=-\epsilon^{ab}\partial_{a}\bar{A}_{b}\,,\ \ \ \ \pi^{a}=-\epsilon^{ab}\left(\partial_{b}\bar{A}_{r}-\partial_{r}\bar{A}_{b}\right)\,.
\ee
Then, equations \eqref{a1}, \eqref{a2} are replaced by 
\begin{eqnarray}
\bar{A}_{a} & = & \bar{F}_{,a}-\gamma^{-\frac{1}{2}}\gamma_{ac}\epsilon^{cb}G_{,b}+\mathcal{O}(r^{-1})\,,\label{bcm1}\\
\bar{A}_{r} & = & \bar{a}_{r}^{(2)}r^{-2}+\mathcal{O}(r^{-3})\,.\label{bcm2}
\end{eqnarray}

It is important to realize that the new potential incorporates with it the \textit{additional} 
variable $\bar{F}$, which was not present in the electric representation and drops out 
from eqs. \eqref{abardef}.

 There are now also magnetic improper
gauge transformations with an associated parameter $\bar{\lambda}_{\left(0\right)}$,
which is independent of the ``electric'' $\lambda_{\left(0\right)}$.
Under a magnetic BMS transformation $\bar{F}$ and $G$ transform
according to
\begin{equation}
\bar{F}\rightarrow\bar{F}+\bar{\lambda}_{\left(0\right)}\,,\label{eq:F-1}
\end{equation}
\begin{equation}
G\rightarrow G\,.\label{eq:G-1}
\end{equation}
The electric and magnetic radial momenta $\pi^r$, $\bar\pi^r$, are related  $G$ and $\bar G$ through,
\be{piG}
\pi^r= \gamma^{\frac{1}{2}}\nabla^2 G, \ \ \ \ \ \ \bar\pi^r= \gamma^{\frac{1}{2}}\nabla^2 \bar G.
\ee

If one demands that $G$ and $\bar{G}$ be regular on the sphere,
there is no room for a zero mode in
the electric and magnetic BMS charges.  The zero modes must be introduced through Dirac string singularities. 

For a magnetic pole of strength $g$ at the origin, on has 
\begin{equation}
A_{\phi}=g(1-\cos\vartheta)\,,\label{Apole}
\end{equation}
\begin{equation}
\bar{G}=g\log(1+\cos\vartheta)\, .\label{Gpole}
\end{equation}
For an electric pole of strength $e$, which in the electric representation has
\begin{equation}
\pi_{(0)}^{r}=\gamma^{\frac{1}{2}}e\,,\label{qpole}
\end{equation}
one now writes 
\begin{equation}
\bar A_{\phi}=e(1-\cos\vartheta)\,,\label{epole}
\end{equation}
\begin{equation}
G=e\log(1+\cos\vartheta)\, .\label{egpole}
\end{equation}
If one admits Dirac string singularities in $G$ and $\bar G$ one must also do so for $F$ and
 $\bar{F}$ in order, for example, to be able to implement rotations. This is so because under
  a rotation the monopole potentials change by a singular gauge transformation.
  
  \subsubsection{Electric-magnetic duality invariant notation}
  
  It is useful to introduce a compact notation that makes electric-magnetic duality
   invariance of the theory manifest. This is achieved by writing 
   \be{AM}
A_a^M=\partial_a F^M +\epsilon^{M}_{\ \ N}\star\nabla_a G^N,
\ee
\be{twopotnot}
 A^{M}=\begin{pmatrix}A\\
\bar{A}
\end{pmatrix}, \  \ 
 N^{M}=\begin{pmatrix}N\\
\bar{N}
\end{pmatrix}, 
\ee
\be{twopotnot2}
F^{M}=\begin{pmatrix}F\\
\bar{F}
\end{pmatrix},   \  \ 
G^{M}=\begin{pmatrix}G\\
\bar{G}
\end{pmatrix},
\ee
where
\be{qm}
Q^M=\left.\frac{}{}
\gamma^{\frac{1}{2}}\nabla^2 G^M \right|^{+\infty}_{-\infty},
\ee
are the electric and magnetic charges.    
\subsection{Time translations: Improved generator}

\label{imp}

\subsubsection{Analysis starting from the electric representation}

Rather than employing the electric-magnetic invariant formalism ab initio, we prefer to start from the ``electric" representation and then use elements of duality to ``patch it" in order to cast final results in a duality invariant form. This we do for expediency, but -- more importantly -- because in the case of gravitation, where the full asymptotic duality invariant formalism has not yet been developed, one can still perform the same steps, starting from the available electric representation. 

It will suffice to analyze time translations. Spatial translations are taken care of in the same manner with a bit more of algebra. This is done in \cite{book}.

If one considers the Hamiltonian for a motion corresponding to a time 
translation, the surface term in the variation of the 
Hamiltonian \eqref{H0}
is given, in the electric representation, by 
\be{dHzero}
\delta H_0=-\left. \oint h^{a}\delta a_{a}^{\left(0\right)}\right|^{\infty}_{-\infty}  \ .
\ee
Equation \eqref{dHzero} may be rewritten separating the electric memory and magnetic charge variations as,
\begin{align}
\delta H_{0} =&-\oint\left.\gamma^{\frac{1}{2}}\nabla^a N\nabla_a \delta F\right|_{-\infty}^{+\infty}  \nonumber \\
& -\oint\left.\gamma^{\frac{1}{2}}\nabla^a \bar N\nabla_a \delta \bar G\right|_{-\infty}^{+\infty}.
\label{dHO}
\end{align}

\subsubsection{Magnetic fiber memory brought in}

If  the parity conditions are used, the first term on the right hand side on \eqref{dHO} vanishes, but the second does not. It may be written as,
\be{deltaH}
\oint \bar N \delta \bar Q. 
\ee

Equation \eqref{deltaH} shows that in order to improve $H_0$ one must add to it 
a term proportional to the \textit{magnetic} gauge constraint\footnote{The magnetic gauge generator, $-\bar\pi^i_{,i}$, can be treated properly by keeping in Dirac's ``total Hamiltonian" the full constraint $\vec\pi_{mag}=0$ and $\vec\pi_{el}+\nabla\times\vec {\bar A}=0$, whose curl is second class, while their divergence $\nabla\cdot\vec\pi_{mag}$ is first class. The details of that treatment will not be needed herein.},
\be{impterm}
\int-\bar{\pi}_{,i}^{i}\bar N d^{3}x.
\ee
This means that it is essential to bring in the magnetic sector in order to properly define the spacetime 
translation generators. \textit{The improvement cannot be made solely within the 
electric sector}. In other words, a deformation consisting only of a spacetime translation by itself  
does not have a well-defined generator. Only when one adds to it a movement 
along the fiber whose magnitude is $\bar N$, does the generator exist.
 It is this improved generator which deserves to be called 
 $P^0$. Its numerical value is the same as the 
 original $H_{0}$ because the other term \eqref{impterm} vanishes weakly.
  
  The need for the addition of the magnetic gauge transformation is simple to understand. {\it It brings in the magnetic fiber memory}, that -- unlike the magnetic charge -- is not present in the purely electric formulation, because only the gauge invariant curl of the magnetic potential appears in it.
  
  The magnetic analog of \eqref{Fdott} is
  \be{FbarN}
  \dot{\bar F}=\bar N.
  \ee
  
  It is remarkable how, guided just by the need to have a well defined Hamiltonian, one is compelled to bring in the magnetic sector in full force\footnote{One could have
  tried to stay within the electric sector by demanding that the magnetic charge $\bar{Q}$ should be a passive spectator given as an
  ``external field'', and not varied in the action principle. For consistency it should be given so that $\dot{\bar{Q}}=0$
  (Eq. \eqref{dQ0}) up to a Lorentz transformation. But the boundary term in \eqref{dHO} would not vanish if $\delta \bar G=\bar G_{,a}\xi^a$, so this possibility is not tenable if one wants to have Lorentz invariance. Thus it is ultimately Lorentz invariance which forces one to bring in the magnetic sector with its own independent life.}.
  
  Had we have started from the magnetic sector we would have obtained an equation identical to \eqref{dHO} but with the electric and magnetic roles reversed. After duality invariance is fully implemented, the variation of the improved generator of time translations will read,
\be{H0dinvdual}
\delta H_{0} \approx -\left. \oint\gamma^{\frac{1}{2}}\nabla^a N^{M}\delta (\nabla_a F_M)\right|_{-\infty}^{+\infty}.
\ee
Here the weak equality means that terms proportional to the electric and magnetic constraints $-\pi^i_i$ and $-\bar \pi^i_i$ have been dropped.

The variation \eqref{H0dinvdual} vanishes when the parity conditions hold, but it will be useful to know its form even when they do not, when we evaluate emission and absorption rates in section \ref{ssrates}.

\subsection{Lorentz generators. Spin from charge}

We again start from the electric representation and at the end cast the results in a manifestly duality invariant form.

We have
\be{Hlorentz}
H_0^{\mbox{\tiny Lorentz}}=-\int d^3 x \xi^i F_{ij}\pi^j,
\ee
where $\xi$ are the Lorentz Killing vectors.
The surface term in its variation reads
\be{surflor}
\delta H_0^{\mbox{\tiny Lorentz}}=-\oint  \xi^{a}\delta 
A_{a}\pi^{r}\Big|^{+\infty}_{-\infty}.
\ee
To improve the generator $\mH_0$ we add an electric gauge generator,
but this time \textit{with the surface term included}, namely, 
\begin{eqnarray}
\mathcal{G}^{\mbox{\tiny Lorentz}}&=&\int-\pi_{,i}^{i}\lambda_{\mbox{\tiny Lorentz}}\,d^{3}x + \\&+& \left.\oint\lambda_{\mbox{\tiny Lorentz}}\left(\infty\right)\pi^{r}\right|_{-\infty}^{+\infty},
\label{GLorentz}
\end{eqnarray}
with 
\be{llorentz2}
\lambda_{\mbox{\tiny Lorentz}}(\infty)=\xi^{a}_{\mbox{\tiny Lorentz}}A_a.
\ee
  One then finds that the variation of 
 \be{Himplor}
{H}_{\mbox{\tiny improved}}^{\mbox{\tiny Lorentz}}= H_0^{\mbox{\tiny Lorentz}}+\mathcal{G}^{\mbox{\tiny 
Lorentz}},
 \ee
 does not have a surface integral. 
 
 \subsubsection{Lie derivative restored}
 
 The improvement of the Lorentz generator $\mH_0^{\mbox{\tiny Lorentz}}$ has an important 
 geometrical consequence, in that it restores the Lie derivative at infinity. Indeed, the change in
  $A_{i}$ given by the generator $\mH_0^{\mbox{\tiny Lorentz}}$ is given by 
 \[
 \delta_{0}A_{i}=\xi^{j}F_{ji}=\mathcal{L}_{\xi}A_{i}-\partial_{i}\left(\xi^{j}A_{j}\right),
 \]
 so that 
 \[
 \delta_{\mbox{\tiny 
 improved}}A_{a}\left(\infty\right)=\mathcal{L}_{\xi}A_{a}\left(\infty\right).
 \]
 Therefore, $\mathcal{H}_{\mbox{\tiny improved}}^{\mbox{\tiny Lorentz}}$ is the 
 generator that will correctly implement the symmetry algebra given in section \ref{symalgebra} below.

 \subsubsection{Spin from charge}
 
 The numerical value of the generator \eqref{Himplor} which realizes the improvement of the 
 Lorentz generator is not zero, but it is equal to the surface integral that 
 appears in it. Therefore the numerical value of the angular momentum is not 
 just the volume integral \eqref{Hlorentz}, but it includes a contribution 
\be{lorimpgen}
\left.\oint \xi^{a} A_{a}\pi^{r}\right|_{-\infty}^{+\infty}= \left.\oint \pi^{r} \xi^{a} \partial_{a}F\right|_{-\infty}^{+\infty} + S,
\ee 
where
\begin{eqnarray}
S &=& \left.\oint  \xi^{a}\star\nabla_a \bar G \gamma^{\frac{1}{2}}\nabla^2 G \right|_{-\infty}^{+\infty} \nonumber \\ 
&=&-\frac{1}{2}\left.\oint  \xi^{a}\star\nabla_a G^M\epsilon_{MN} \gamma^{\frac{1}{2}}\nabla^2 G^N \right|_{-\infty}^{+\infty} \!\!\!\!\! . \label{stt}
 \end{eqnarray}
This phenomenon is similar
to the modification of the angular momentum which appears in the presence
of a magnetic pole in abelian and non-abelian gauge theories.The
novelty here is that it occurs already without a magnetic pole.

The spin from charge phenomenon does not happen for energy and momentum
because no surface term analogous to the one appearing in \eqref{lorimpgen}
is included in the translation charge.

\subsubsection{Duality invariant Lorentz generator}
\label{dilg}

The improved electric Lorentz generator,
\be{electricJ}
H_{el}[\xi]=\left. H_0[\xi]+S+ \oint \pi^r \xi^a\partial_a F\right|_{-\infty}^{+\infty},
\ee
is not electric-magnetic duality invariant because, whereas $H_0$ and $S$ have that property, the term proportional to $\pi^r$ does not. Just as it was discussed for translations, it is evident that the appropriate expression is
\begin{eqnarray}
	 H[\xi]&=& \left. H_{el}[\xi] + \oint \bar\pi^r \xi^a\partial_a \bar F\right|_{-\infty}^{+\infty} \nonumber \\
&=&\left. H_0[\xi]+S+ \oint \pi^r_M \xi^a\partial_a F^M\right|_{-\infty}^{+\infty}\nonumber \\
&=& H_0[\xi]+S+ \oint Q_M \xi^a\partial_a F^M. \label{lorentzQ}
\end{eqnarray}
One may think of $\oint\xi^a Q^M F_{M,a}$ as the generator of Lorentz transformations at infinity, and $H_0+S$ as the ``bulk part" (although $S$ is a surface integral).

It will be shown below, in Sec. \ref{angmom}, that the duality invariant angular momentum is conserved (the electric part \eqref{electricJ} is not!).  Since this has been an issue in the literature (in the case of gravitation, which will follow the same lines) it is worth some comment.

First of all one realizes that under improper electric and magnetic gauge transformation, with parameter $\lambda^M_{(0)}=\epsilon^M$, the Lorentz generator changes as,
\be{Lortran}
H[\xi] \longrightarrow H[\xi]  - \oint \epsilon^M \partial_a(\xi^a Q_M),
\ee 
and the new angular momentum is also conserved because $Q_M$ is.

This is just as it happens if one changes the origin for orbital angular momentum, and in our view it is not to be regarded as a difficulty, since the present formalism improper gauge transformations are on the same footing with spacetime translations. All the more so, since a ``pure time translation" carries along with it a rotation along the fiber, due to the fiber memory. Corresponding comments will be given below concerning angular momentum radiation rates.

\subsection{Symmetry algebra}
\label{symalgebra}

The electric and magnetic BMS charges generate improper gauge transformations and therefore commute with the  spacetime translation generators which are invariant under them; and also among themselves.
The action of the BMS charges on the Lorentz generators is given by \eqref{Lortran}. 

Therefore one obtains the algebra
\begin{equation}
\left[Q_M,\mH_{\mbox{\tiny Lorentz}}\right]^*=\ \partial_{a}\left(Q_M\xi_{\mbox{\tiny Lorentz}}^{a}\right)\,,\label{QHcom}
\end{equation}
\begin{equation}
\left[Q_M\left(\vartheta,\varphi\right),P_{\mu}\right]^*=0\, ,\label{eq:QP}
\end{equation}
\begin{equation}
\left[Q_M\left(\vartheta,\varphi\right),Q_N\left(\vartheta',\varphi'\right)\right]^*=0\, ,\label{eq:QP}
\end{equation}
for the charges with the Poincar\'e group and among themselves. The Poincar\'e generators close according to the Poincar\'e algebra.

\subsection{Emission and absorption rates. Charge memory} 
\label{ssrates}

\subsubsection{General formula for emission rates}

Our boundary conditions are appropriate for a closed system, whose Hamiltonian is invariant
 under Poincar\'e and improper gauge transformations, and the corresponding conservation laws hold as a consequence of the fact that as much radiation is coming in as going out.
 
However, the formalism provides expressions for the emission and absorption rates separately. For that purpose one realizes from Eqs. \eqref{dHzero} and \eqref{H0dinvdual} that
\begin{eqnarray}
\dot Q_\alpha &=&\left.
-\oint h^{a}\delta_{\alpha}a_{a}^{\left(0\right)}\right|^{+\infty}_{-\infty} \nonumber \\
&\approx&\left.-\oint \gamma^{\frac{1}{2}}\nabla^a N^M \nabla_a(\delta_{\alpha}F_M)\right|^{+\infty}_{-\infty} .\label{rates} 
\end{eqnarray}
Here $\delta_{\alpha}$ is the variation due to the motion generated by 
the charge $Q_\alpha$.
Thus  $\delta F^M=\epsilon^M$
for gauge transformations, $\delta F^M=N^M$
for time translations, and $\delta F^M= F^M_{\ \  ,a}\xi^a$
for rotations. The purely electric form is incomplete for the magnetic charges and the Lorentz charges, because it misses the effect of the magnetic memory. This is not seen by $P_\mu$ or by the electric BMS charge.

Then, the emission rates are read from the upper endpoint in \eqref{rates} and the absorption rates from the lower one. In this way, one obtains the following results.

\subsubsection{BMS charge}

\be{nnnn}
\dot Q^M=\partial_{a} h^{a}_{M}\Big|^{+\infty}_{-\infty} =\gamma^{\frac{1}{2}}\nabla^2 N^M \Big|^{+\infty}_{-\infty},
\ee
where,
\[
h_{N}^{a}=\begin{pmatrix}h^{a}\\
-\star h^{a}
\end{pmatrix}.
\]
This equation is to be interpreted as giving either $\left[+\dot\pi^r (\infty)\right]$,
 or $\left[-\dot\pi^r (-\infty)\right]$. These are {\it not} to be thought of as the rate of change of
  two different charges, but rather as the rates of change of one and the same charge, due to
   outgoing and incoming radiation respectively; which must be calculated using the two replicas
    of space that form the hourglass. When the parity conditions hold the $Q^M$ are conserved.

On sees from \eqref{nnnn}, in analogy with \eqref{Fdott}, that the BMS charge
 also ``remembers'' the news and that, in this sense, the Laplacian of $N$ is the ``charge memory''.
 We will see in [\citenum{BGP2}], that when a cosmological constant is introduced the fiber and charge memories are
 different and that the fiber memory appears to be more fundamental.

\subsubsection{Energy}

Similarly, one finds for the energy 
\begin{eqnarray}
\frac{dP^{0}}{dt} &=&-\left.\oint \gamma^{-\frac{1}{2}}\gamma_{ab}h^{a}h^{b} \right|^{\infty}_{-\infty} \nonumber
\\ &=& \left.-\oint \gamma^{-\frac{1}{2}}\nabla^a N^M \nabla_a N_M\right|^{\infty}_{-\infty} .\label{dtP} 
\end{eqnarray}

\subsubsection{Angular momentum}
\label{angmom}
The equations for the rate of
 change of the BMS charges and the energy given above can be
 expressed solely in terms of quantities
  defined in the electric sector. This is not the case for the angular momentum which as argued before,
   needs the magnetic sector for its very definition. Therefore,  the rate can be read only from the second
 term on the right hand side of  Eq. \eqref{rates} , which yields,  
\begin{equation}
\frac{d\vec{J}}{dt}=\left.-\oint \gamma^{\frac{1}{2}} \nabla^a N^M\nabla_a\left(\L_{\vec{\xi}} F_{M}\right) \right|^{\infty}_{-\infty}\,\label{dtJ},
\end{equation}
an expression that can be rewritten, with the help of \eqref{nnnn}, as,
\begin{equation}
\frac{d\vec{J}}{dt}=\oint \dot Q_M  F^M_{\ ,a}\vec \xi^a.\label{dtJQ}
\end{equation}

The last expression shows that when the parity conditions hold, so that $\dot Q^M=0$, the angular momentum, Eq. \eqref{lorentzQ}, is conserved, as it was announced and discussed in Sec \ref{dilg}.

 Note that Eq. \eqref{dtJ} involves the 
   variable $\bar F$ which does not appear in the electric sector. This is a consequence, in turn, 
   of the fact that the angular momentum changes under the action of the magnetic BMS charge.

The interpretation of these equations is that the left hand sides are the rate of change of one and the
 same energy and angular momentum due to outgoing and incoming radiation. Therefore, the volume 
 integrals appearing in the definition of $P_0$ and $J$ (see Eq. \eqref{H0}), are to be thought of as
  evaluated on the upper half of the hourglass in the calculation of outgoing radiation and on the lower
   half in the calculation of incoming radiation. One does not integrate over the whole hourglass
    because this would lead to the same overcounting  encountered for the electromagnetic 
    charges. 
    
    Just as it was the case with the angular momentum itself, the physical cogency of Eq. \eqref{dtJ} giving its rate of change, deserves a brief comment. The time rate of change of $F^M$ is invariant under (improper) gauge transformations. If one agrees to keep the gauge frame fixed, that is, if one only moves in the course of time on the fiber as dictated by the fiber memory, then $F_M(t)$ is determined by the equations of motion -- in a gauge invariant manner - once $F_M(t=0)$ is given. This means that if one were absorbing angular momentum at infinity so as to, say, make a top start spinning, then one would in principle be able to determine $F_M(t=0)$ and thus learn how the BMS origin in \eqref{dtJ} is shifted from the one arbitrarily chosen on the fiber.

\section{Gravitational field}

\label{sec:Gravfield}

\subsection{Correspondence with electromagnetism}

\label{subsec:corresp}

In this section we analyze the gravitational field along the same
lines that we analyzed above the electromagnetic field. The parallel
between both cases is so close that it permits to make the following
discussion succinct. The correspondence is as follows: The $\ell=0$
mode of the improper gauge symmetry generated by the total electric
charge $Q$ is the analog of the $\ell=0$, $\ell=1$ modes of the
Bondi-van der Burg-Metzner-Sachs supertranslation, which are the ordinary
translations generated by $P_{\mu}$. The modes with $\ell\geq1$
of the improper gauge symmetry correspond to the modes $\ell\geq2$
of the supertranslations. Therefore, altogether, one has the correspondence:
\[
\underbrace{\left(Q\left(\vartheta,\varphi\right),P_{\mu}\right)}_{\text{electromagnetism}}\longleftrightarrow\underbrace{{\cal P}\left(\vartheta,\varphi\right)}_{\text{gravitation}}\,.
\]
On the other hand, the Lorentz transformations play along side: 
\[
\underbrace{J_{\mu\nu}}_{\text{electromagnetism}}\longleftrightarrow\underbrace{J_{\mu\nu}}_{\text{gravitation}}\,.
\]
There is, as emphasized before, the difference that in the gravitational
case all the generators are given by surface integrals, whereas in
the electromagnetic one since the background was fixed, the spacetime
translations and the Lorentz transformations were not. But this is
just a technical point which is easily accounted for and does not
hinder at all the close correspondence between both cases. 

The important concept of ``news'' is also present here of course,
since it is the context in which it was originally introduced by Bondi
 \cite{Bondi:1960jsa}. The only difference is that now it is a symmetric
traceless tensor $h^{ab}$, appropriate to describe a gravitational
wave, rather than the vector $h^{a}$ appropriate for an electromagnetic
one. Thus, one has the correspondence:
\[
\underbrace{h_{a}}_{\text{electromagnetism}}\longleftrightarrow\underbrace{h_{ab}}_{\text{gravitation}}\,.
\]
Keeping this in mind, we will essentially write the corresponding
equation without much discussion, because one may translate to gravitation
word by word in each case the corresponding comments from electromagnetism.

\subsection{Asymptotic boundary conditions}

For the gravitational field the canonical variables are the spatial metric $g_{ij}$ and their 
conjugate $\pi^{ij}$. The generators
of surface deformation are given by,

\begin{align*}
\mathcal{H}_{\perp} & =\frac{2}{\sqrt{g}}\left(\pi^{ij}\pi_{ij}-\frac{1}{2}\pi^{2}\right)-\frac{1}{2}\sqrt{g}{}^{\left(3\right)}R\approx0\,,\\
\mathcal{H}_{i} & =-2\pi_{i\ |j}^{\ j}\,\approx0.
\end{align*}
Here we have set the cosmological constant equal to zero, and have chosen units such that $8\pi G=1$. 
The deformation parameters that multiply $\mathcal{H}_{\perp}$ and $\mathcal{H}_{i} $ 
in the Hamiltonian are the lapse $N^{\perp}$ and the shift $N^{i}$.

\subsubsection{Power expansion at large distances}

Since our spacelike surfaces are asymptotically null, we must take
as a starting point a coordinate system for the Schwarzschild metric
which incorporates this property. This is provided by the
Eddington-Finkelstein coordinates in terms of which the line element
reads, 
\begin{eqnarray}
ds^{2} =-(dx^{0})^{2}&+&dr^{2}+r^{2}\left(d\theta^{2}+\sin^{2}\theta d\phi^{2}\right)+ \nonumber  \\
&+& \frac{M}{4\pi r}\left(dx^{0} -  dr\right)^{2} .
\label{metricgc}
\end{eqnarray}

The next step is to pass to hyperbolic coordinates, through the change of variables \eqref{emb1}-\eqref{emb2},
extract the asymptotic
form of the resulting expression, and proceed by trial and error.

The resulting boundary conditions, in the form of a power law expansion at large distances are given in \cite{book}. We only need to know for the present purposes, that the most general deformation that preserves them is parameterized by a function 
$\epsilon_{\left(1\right)}^{\perp}$ (infinitesimal supertranslation) and two vectors, $\vec \omega$ (infinitesimal rotation) and $\vec \beta$ (infinitesimal boost). The analogs of the asymptotic parts  $a_a^{(0)}$  of the vector potential $A_a$ and of the news $h_a$ are now symmetric traceless tensors $\tilde f^{ab}_{(1)}$ and $h_{ab}$, respectively. They are build out of the leading and subleading terms in the power expansions of $g_{ij}$ and $\pi^{ij}$.

\subsubsection{Parity conditions}

In addition to the power law decays it is necessary to introduce parity conditions. This is achieved by splitting $\tilde f^{ab}_{(1)}$ and $h_{ab}$ in longitudinal and transverse parts as follows,
\begin{equation}
\tau_{0}\tilde{f}_{ab}^{\left(1\right)}=\nabla_{ab}F+\star\nabla_{ab}\bar{G}\,,\label{f1ab}
\end{equation}
\begin{equation}
h_{ab}=\frac{1}{2}\left(\nabla_{ab}N+\star\nabla_{ab}\bar{N}\right) \,\label{habdef},
\end{equation}
These equations correspond to \eqref{eq:divcurl1}--\eqref{eq:divcurl2} in electromagnetism.

In the above equations the operators $\nabla_{ab}$ and $\star\nabla_{ab}$, given by
\begin{eqnarray}
\nabla_{ab}&=&2(\nabla_{a}\nabla_{b}+\nabla_{b}\nabla_{a}-\gamma_{ab}\nabla^{2}),\\ \ \star\nabla_{ab}&=&2\sqrt{\gamma}\gamma^{cd}\left(\epsilon_{ac}\nabla_{b}\nabla_{d}+\epsilon_{bc}\nabla_{a}\nabla_{d}\right)\,,\label{nablas}
\end{eqnarray}
are the tensor analogs of the vector gradient, $\nabla_{a}$, and
curl $\star\nabla_{a}$ appearing in \eqref{eq:divcurl1} and \eqref{eq:divcurl2}. 

These operators
were used by Regge and Wheeler in their analysis of the stability
of a Schwarzschild singularity  \cite{Regge:1957td}, and obey the
key properties
\begin{equation}
 \nabla^{ab}\left(\star\nabla_{ab}\right)=\star\nabla_{ab}\left(\nabla^{ab}\right)=0,
 \end{equation}
 when they act on scalar functions, just as their vector counterparts.
Their kernel is spanned by the $\ell=0$ and $\ell=1$ modes of the
corresponding scalar functions on which they act.

The parity conditions will be then the following,
\be{parity}
\left. (F, N,\bar N)\right|_{+\infty} =
\left. (F, N,\bar N)\right|_{-\infty} \ \ 
\mbox{for each } (\vartheta,\varphi),
\ee
in close analogy with Eq. \eqref{parity} for electromagnetism.

\subsubsection{Supertranslation memory}

Consider a time translation:
\begin{equation}
  \epsilon_{\left(1\right)}^{\perp}=1.
\end{equation} 
Einstein's equations in Hamiltonian form then yield,
\begin{equation}
  \dot{\tilde{f}}_{ab}^{(1)}=2h_{ab}, \label{dotfab}
\end{equation}
which implies 
\begin{equation}
  \dot{F}=N. \label{cien}
\end{equation}
Therefore, when time $\delta t$ elapses a supertranslation of magnitude
\begin{equation}
  \delta F=N \delta t, \label{FNt}
\end{equation}
takes place. This is the supertranslation memory effect, analogous to the fiber 
memory of electromagnetism discussed in section \ref{sec:Poinc}.

\subsection{Electric and magnetic BMS charges}

We saw in the electromagnetic case that it was necessary to employ, asymptotically on the hourglass an 
electric-magnetic duality invariant formalism, in order to be able to improve the
generators. The same 
will occur in gravitation. In that case we do not possess at the moment an explicit 
electric-magnetic duality invariant description of the linearized theory on the hourglass, which is what is needed at large distances. However, it is reasonable to 
assume that such a description exists, and that it can be constructed along lines similar to those employed succesfully for asymptotic 
planes  in [\citenum{Henneaux:2004jw, Bunster:2006rt}].

Fortunately, it turns out that assuming the existence of the asymptotic electric-magnetic duality invariant description, one 
can conjecture by analogy some of the elements that are needed. The coherence of the results thus obtained reinforces the hypothesized  
existence of the electric-magnetic representation. We now pass to discuss those elements.

\subsubsection{Electric BMS charge}

If one varies the Hamiltonian  
\begin{equation}
H_{0}=\int d^{3}x\left(N^{\perp}\mathcal{H}_{\perp}+N^{i}\mathcal{H}_{i}\right)\,,
\end{equation}
in the electric representation, with the Lorentz parameters $\vec \omega$, $\vec\beta$ set equal to zero,
one finds
\begin{equation}
\delta H_{0}=-\oint \epsilon_{\left(1\right)}^{\perp}\left.\left[\delta\mathcal{P}+\frac{1}{2}h^{ab}\delta\left(\tau_{0}f_{ab}^{\left(1\right)}\right)\right]\right|_{-\infty}^{+\infty}.
\label{eq:delta}
\end{equation}
The explicit expression for $\mathcal{P}(\vartheta,\varphi)$ is given in \cite{book}. Equation \eqref{eq:delta} identifies $Q\left(\vartheta,\varphi\right)$
 \be{Qgrav}
 Q=\left. \mathcal{P}\right|^{+\infty}_{-\infty},
 \ee
 as the (electric) supertranslation charge. In the analogy with 
 electromagnetism, the $l=0$ and $l=1$ of the charge \eqref{Qgrav} correspond
  to spacetime 
 translations whereas those with $l\geq 2$ correspond to the electromagnetic 
 charges with spherical modes $l\geq 1$.
The first term on the right hand side of \eqref{eq:delta} may be compensated in the 
standard manner by defining a partially improved Hamiltonian $\tilde{H}_{0}^{\text{\tiny{elec}}}$ through
\be{H0tilde}
\tilde{H}_{0}^{\text{\tiny{elec}}}=H_{0}+\oint \epsilon_{\left(1\right)}^{\perp}\left(\vartheta,\varphi\right)Q\left(\vartheta,\varphi\right).
\ee
The Hamiltonian \eqref{H0tilde} is the analog of the Maxwell electric Hamiltonian for spacetime 
translations {\it and} improper gauge transformations, and just as that one it will 
need to be improved to eliminate the surface term 
\begin{align}
\oint \frac{1}{2}\epsilon_{\left(1\right)}^{\perp}h^{ab}\delta\left(\tau_{0}\tilde f_{ab}^{\left(1\right)}\right) \Big|_{-\infty}^{+\infty}&= \label{niexp}\\
\oint\frac{1}{4}\gamma^{\frac{1}{2}}\epsilon_{\left(1\right)}^{\perp}\left(\nabla_{ab}N +\star\nabla_{ab}\bar{N}\right)&\left(\nabla_{ab}\delta 
F+\star\nabla_{ab}\delta\bar{G}\right)\Big|_{-\infty}^{+\infty}. \nonumber 
\end{align}
The term proportional to $\delta F$ on the right hand side of Eq. \eqref{niexp} vanishes 
when the parity conditions hold but the one proportional to $\delta \bar{G}$, which reads
\be{dggr}
\delta H_0=-\oint \bar\eta \delta \bar G\Big|_{-\infty}^{+\infty},
\ee
with
\be{eta}
\bar\eta = \frac{1}{4}\nabla^{ab}\left[\epsilon^\perp_{(1)}\left(\nabla_{ab}\bar N - \star\nabla_{ab}N\right)\right]
\ee
does not.

\subsubsection{Magnetic BMS charges}

In order to eliminate \eqref{dggr} one should supplement the Hamiltonian acting with the generator of magnetic BMS transformations, whose form we do not know, but which should be such that the surface term in its variation should read
\be{stv}
-\oint\bar\epsilon^\perp_{(1)}\delta\bar{\cal P} \Big|_{-\infty}^{+\infty} .
\ee
Here, by definition, $\bar{\cal P}$ is the magnetic supertranslation charge and $\bar\epsilon^\perp_{(1)}$ is the magnetic deformation parameter.

 So we must have,
$$
\oint\bar\epsilon^\perp_{(1)}\delta\bar{\cal P} = \oint \bar\eta \delta \bar G,
$$
and the question is: what is the relationship between $\bar{\cal P}$ and $\bar G$?.

This can be established by recalling from electromagnetism that one would like the parameter $\eta$ to bring the magnetic memory. So we set
\be{epsone}
\epsilon^\perp_{(1)}=1,
\ee
in which case the boundary term reads
\begin{eqnarray}
	\delta H_0 &=& -\oint \nabla^4\bar N\delta (\gamma^{\frac{1}{2}}\bar G)\Big|_{-\infty}^{+\infty} \nonumber \\
&=& -\oint \bar N\delta \left( \nabla^4(\gamma^{\frac{1}{2}}\bar G)\right)\Big|_{-\infty}^{+\infty}, \label{bt2}
\end{eqnarray}
where
\begin{eqnarray}
\nabla^{4}&=&\nabla_{ab}\nabla^{ab}
=\star\nabla_{ab}\star\nabla^{ab} \nonumber \\ &=& 8\left[\left(\nabla^{2}\right)^{2}+2\nabla^{2}\right]. \label{nabla4}
\end{eqnarray}
Comparison with the magnetic analog of Eq. \eqref{cien} then gives
\be{pbar}
\bar{\cal P}= \gamma^{\frac{1}{2}}\nabla^4 \bar G
\ee
The identification \eqref{pbar} will have a significant consistency check when we discuss angular momentum below.

In order to account for the $l=0$ and $l=1$ modes of $\bar{\cal P}$ (magnetic translations) one would need to bring in Dirac strings into $\bar G$ because those 
modes are in the kernel of $\nabla^4$.

\subsection{Lorentz generators}

If one works solely in the electric representation one 
finds that if one considers the motion corresponding to a Lorentz transformation, 
with infinitesimal rotation and boost parameters $\vec{\omega}$ and 
$\vec{\beta}$, one must improve the Hamiltonian by adding to it the surface term, 
\[
\vec{\omega}\cdot\vec{J}_{\mbox{\tiny el}}+\vec{\beta}\cdot\vec{K}_{\mbox{\tiny el}},
\]
where $\vec{J}_{\mbox{\tiny el}}$ and $\vec{K}_{\mbox{\tiny el}}$ are surface integrals whose expressions are given in \cite{book}.

When this is done the generators are well-defined. No additional surface 
integral containing the news, analogous to \eqref{niexp} appears. This is reasonable because the Lorentz motion lies within the hourglass. 

The electric generators thus obtained are the analog of the electromagnetic 
angular momentum \eqref{electricJ}. 

If one takes $\epsilon^\perp_{(1)}=1$, and evaluates the rate of change of $\vec J_{el}$ , one finds, either by direct calculation from Einstein's equations or, better, by using Eq. \eqref{qdotgrav} below,
\be{djeldt}
\frac{d\vec J_{el}}{dt}=-\oint (\gamma^{\frac{1}{2}}\nabla^4 \bar G)\bar N_{,a} \vec\xi^a\bigg|_{-\infty}^{+\infty} .
\ee
Eq. \eqref{djeldt} is the analog of \eqref{dtJ} for electromagnetism. It provides a consistency check of the definition \eqref{pbar} because if we bring in the magnetic analog  $\bar F$ of $F$, and postulate the magnetic memory equation,
\be{maganalog}
\dot {\bar F}= \bar N,
\ee
then,
\be{Jcons}
\vec J= \vec J_{el}+ \oint \bar {\cal P}\bar F_{,a} \vec\xi^a\bigg|_{-\infty}^{+\infty}
\ee
is conserved
\be{djdtgr}
\frac{d\vec J}{dt}=0,
\ee
when the parity conditions hold.

So, by appealing to electric-magnetic duality one can find a conserved angular momentum in general relativity, even in the presence of radiation, but provided the net radiation flux is zero. 

For boosts one must include an extra term (see comment at the end of the next subsection). Thus one has in general,
\be{hbl}
H_{\mbox{\tiny Lorentz}}= H_{\mbox{\tiny Lorentz}}^{\mbox{\tiny el}}+ \oint  \bar Q\left(\xi^a\partial_aF-\frac{3}{2}\nabla_a\xi^aF \right) .
\ee 
(The second term $\nabla_a\xi^a$ vanishes for rotations).

\subsection{Symmetry algebra}

The analog of \eqref{QHcom} and \eqref{eq:QP} for electromagnetism
is 
\begin{eqnarray}
\left[Q^M\left(\vartheta,\varphi\right),Q^N\left(\vartheta',\varphi'\right)\right]^{\star} &=&0\nonumber \\
\left[Q^M\left(\vartheta,\varphi\right),\vec{J}\right]^{\star} &=&\partial_{a}\left(\vec{\xi}_{\text{R}}^{a}Q^M\left(\vartheta,\varphi\right)\right) \label{BMSalgebra}\nonumber\\
\left[Q^M\left(\vartheta,\varphi\right),\vec{K}\right]^{\star} &=&\partial_{a}\left(\vec{\xi}_{B}^{a}Q^M\left(\vartheta,\varphi\right)\right) \\ &+&\frac{3}{2}Q^M\left(\vartheta,\varphi\right)\left(\nabla_{a}\vec{\xi}_{B}^{a}\right),\nonumber
\end{eqnarray}
while the Lorentz generator $\vec{K}$ and $\vec{J}$ close among themselves in 
the Lorentz algebra.

We have used Dirac brackets $\left[\;,\;\right]^{\star}$ here because,
as explained in the introduction it is only through them that the
surface term alone can act as a generator. If one wanted to use Poisson
brackets one would have to add to the surface term the weakly vanishing
volume part of the generator.

For the electric generators $Q$ and the Lorentz generators, equations \eqref{BMSalgebra} can be obtained directly from the algebra of surface deformations in spacetime \cite{Teitelboim:1972vw,Teitelboim:1973yj} (see \cite{book} for details). It is then extended to the magnetic generators by duality.

\subsection{Emission and absorption rates. Charge memory}

\subsubsection{General formula for emission rates}

In this case we only possess the formula stemming from electric sector, that is,
\be{qdotgrav}
\dot Q_\alpha= \oint \frac{1}{2}h^{ab}\delta_\alpha\left(\tau_{0}\tilde f_{ab}^{\left(1\right)}\right) \Big|_{-\infty}^{+\infty}.
\ee
The analog of the second expression on the right hand side of \eqref{rates} is not obvious to guess because, this time, under a duality transformation, one must turn the electric time  into magnetic time. That is, one would have to compare motions that have $\epsilon^\perp_{(1)}=1$, $\bar  \epsilon^\perp_{(1)}=0$ with those with $\epsilon^\perp_{(1)}=0$, $\bar  \epsilon^\perp_{(1)}=1$ .

By applying \eqref{qdotgrav} one obtains the following results.

\subsubsection{Electric BMS charges}

\begin{align}
\frac{\partial Q}{\partial t}&=\left(-\frac{1}{\sqrt{\gamma}}h^{ab}h_{ab}+\nabla_{a}\nabla_{b}h^{ab}\right)\bigg|^{+\infty}_{-\infty} \,,\nonumber \\
&= -\frac{1}{4}\gamma^{\frac{1}{2}}\bigg((\nabla^{ab}N)(\nabla_{ab}N) + (\nabla^{ab}\bar N)(\nabla_{ab}\bar N) + \nonumber \\ &+ 2(\nabla^{ab}N)(\star\nabla_{ab}\bar N) +
+\gamma^{\frac{1}{2}}\nabla^4 N\bigg)\bigg|^{+\infty}_{-\infty}, \label{Aeq}
\end{align}
(This equation,  as well as its relationship with the emission rate, were found previously in \cite{Barnich:2009se,Barnich:2011mi}; but they interpreted it as meaning that the Hamiltonian cannot be improved if
   $h_{ab}\neq 0$.)

\subsubsection{Magnetic BMS charge}

The (electric) time derivative of the magnetic BMS charge cannot be obtained from the purely electric sector formula \eqref{qdotgrav}, although $\bar {\cal P}$ does appear in the electrir sector. One must resort to its definition \eqref{pbar} and to the equation of motion \eqref{dotfab}. This yields,
\be{Beq}
\frac{\partial\bar Q}{\partial t}=\gamma^{\frac{1}{2}}\nabla^4
 \bar N\Big|^{+\infty}_{-\infty}
\ee

Note that there is no symmetry between the rates of change of $Q$ and $\bar Q$. That is quite alright because one should not expect any: the duality counterpart of \eqref{Aeq} should be the rate of change of $\bar Q$ with respect to a {\it magnetic} time displacement with $\bar\epsilon^\perp_{(1)}=1$, and  $\epsilon^\perp_{(1)}=0$.

In the same vein, one could define a variable $G$ thorough
\be{calP}
\gamma^{\frac{1}{2}}\nabla^4  G = \cal P.
\ee
Then the derivative of $G$ with respect to electric time would not be equal to $N$ as one can see from \eqref{Aeq}. However, one would expect its derivative with respect to magnetic time to be given by $\bar N$, in analogy with \eqref{Beq}.

\subsubsection{Angular momentum}

One may write, in analogy with \eqref{dtJQ} in electromagnetism

\begin{equation}
\frac{d\vec{J}}{dt}=\oint \dot {\cal P}_M F^M_{\ ,a}\vec \xi^a \label{dJdtgrav}
\end{equation}
where $\dot {\cal P}$ and $\dot{\bar {\cal P}}$ are given by \eqref{Aeq} and \eqref{Beq}.

It should be stressed that formula \eqref{dJdtgrav} has not been proven, but just conjectured by analogy and ``informed guess".
Only the conservation of  $\vec J$ when $\dot {\cal P}$ and $\dot{\bar {\cal P}}$ vanish has been proven (once \eqref{maganalog} has been postulated!). This is because, in the lack of a complete asymptotic two potential theory, we do not posses an analog of the second expression on the right hand side of \eqref{rates}.  But the presumption is  that \eqref{dJdtgrav} will survive the complete development of the asymptotically duality invariant description.

All the comments made for the electromagnetic case in connection with the angular momentum and with its rate of change apply here as well.
   
\subsection{Taub-NUT and Kerr solutions}

To conclude we consider two fundamental solutions of Einstein's equations for which it is important to verify that they fit into the present treatment. Especially so because of their relationship with magnetic charge and with angular momentum, concepts which have been of central interest throughout this work. They are Taub-NUT space and the Kerr solution respectively. 

One can verify (see \cite{book}) that this two solutions can be brought by a change of coordinates to comply to our boundary conditions. One finds that for Taub-NUT,
\begin{equation}
F=0\,,\label{eq:F-2}
\end{equation}
 and 
\begin{equation}
\bar{G}=-N\log\left(1+\cos\vartheta\right)\, ,\label{G}
\end{equation}
which  are exactly the expressions \eqref{Gpole} of electromagnetism
for a magnetic pole of charge $g=-N$, with the Dirac string going
through the south pole\footnote{When one discusses Taub-NUT on surfaces which are asymptotically planes,
as it was done in \cite{Bunster:2006rt}, one finds, that in order
to satisfy the Regge-Teitelboim boundary conditions one must take half of the string to come out
of the south pole and the other half to come out from the north pole.
No such requirement is present here, where one can take just one string
going out through any point on the sphere. }. Here $N$ is the Taub-NUT parameter.

For the Kerr metric, one finds that the value of the energy is given by
\be{P0Kerr}
P^{0}  =\oint {\cal P}=M,
\ee
and that of the angular momentum \eqref{Jcons} by
\be{KerrJ}
J_{z} =aM\,.
\ee
Only the electric part of \eqref{Jcons} contributes to  \eqref{KerrJ} becuase the magnetic part vanishes. $M$ and $a$ are the standard mass and specific angular momentum parameters of the Kerr solution.

\section*{Acknowledgments}

%%%%%%%%%%%%%%%%%%%%%%%%%%%%%%%%%%%%%%%%%%%%%%%
We express our gratitude to Professor Anna Ceresole for her continued encouragement throughout this work.
The Centro de Estudios Cient\'ificos (CECs) is funded by the Chilean
Government through the Centers of Excellence Base Financing Program
of Conicyt. C.B. wishes to thank the Alexander von Humboldt Foundation
for a Humboldt Research Award. The work of A.P. is partially funded
by Fondecyt Grants N\textsuperscript{o} 1171162 and N\textsuperscript{o} 1181496.


\begin{thebibliography}{10}

\bibitem{Bondi:1962px}
H.~Bondi, M.G.J. van~der Burg, and A.W.K. Metzner.
\newblock {Gravitational waves in General Relativity. 7. Waves from
  axisymmetric isolated systems}.
\newblock {\em Proc.Roy.Soc.Lond.}, A269:21--52, 1962.

\bibitem{Sachs:1962zza}
R.~Sachs.
\newblock {Asymptotic symmetries in gravitational theory}.
\newblock {\em Phys.Rev.}, 128:2851--2864, 1962.

\bibitem{Sachs:1962wk}
R.~K. Sachs.
\newblock {Gravitational waves in General Relativity. 8. Waves in
  asymptotically flat space-times}.
\newblock {\em Proc. Roy. Soc. Lond.}, A270:103--126, 1962.


\bibitem{Barnich:2001jy}
G.~Barnich and F.~Brandt.
\newblock {Covariant theory of asymptotic symmetries, conservation laws and
  central charges}.
\newblock {\em Nucl. Phys.}, B633:3--82, 2002, hep-th/0111246.

\bibitem{Barnich:2009se}
G.~Barnich and C.~Troessaert.
\newblock {Symmetries of asymptotically flat 4 dimensional spacetimes at null
  infinity revisited}.
\newblock {\em Phys.Rev.Lett.}, 105:111103, 2010, 0909.2617.

\bibitem{Barnich:2011mi}
G.~Barnich and C.~Troessaert.
\newblock {BMS charge algebra}.
\newblock {\em JHEP}, 12:105, 2011, 1106.0213.


\bibitem{Strominger:2013jfa}
A.~Strominger.
\newblock {On BMS Invariance of gravitational scattering}.
\newblock {\em JHEP}, 07:152, 2014, 1312.2229.

\bibitem{Strominger:2017zoo}
A.~Strominger.
\newblock {Lectures on the infrared structure of gravity and gauge theory}
\newblock 2017, 1703.05448, and references therein.

\bibitem{book}
C.~Bunster, A.~Gomberoff and A.~P\'erez.
\newblock {Regge-Teitelboim analysis of the symmetries of electromagnetic and gravitational fields on
 asymptotically null spacelike surfaces,}
 \newblock 1805.03728.
 \newblock {To appear in the forthcoming volume ``Tullio Regge: an eclectic genius, from quantum gravity to computer play,'' Eds. L. Castellani, A. Ceresola, R. D'Auria and P. Fr\'e (World Scientific).}

\bibitem{Regge:1974zd}
T.~Regge and C.~Teitelboim.
\newblock {Role of surface integrals in the Hamiltonian formulation of General Relativity}.
\newblock {\em Annals Phys.}, 88:286, 1974.

\bibitem{Ashtekar:1978zz}
A.~Ashtekar and R.~O. Hansen.
\newblock {A unified treatment of null and spatial infinity in general
  relativity. I - Universal structure, asymptotic symmetries, and conserved
  quantities at spatial infinity}.
\newblock {\em J. Math. Phys.}, 19:1542--1566, 1978.


\bibitem{0264-9381-9-4-019}
A.~Ashtekar and J.D. Romano.
\newblock Spatial infinity as a boundary of spacetime.
\newblock {\em Classical and Quantum Gravity}, 9(4):1069, 1992.

\bibitem{Campiglia:2015qka}
M.~Campiglia and A.~Laddha.
\newblock {Asymptotic symmetries of QED and Weinberg's soft photon theorem}.
\newblock {\em JHEP}, 07:115, 2015, 1505.05346.

\bibitem{Troessaert:2017jcm}
C.~Troessaert.
\newblock {The BMS4 algebra at spatial infinity}.
\newblock 2017, 1704.06223.

\bibitem{Rohrlich}
F.~Rohrlich.
\newblock {\it Classical charged particles}.
\newblock {\em Addison-Wesley, Boston}, 1965.

\bibitem{Teitelboim:1970mw}
C.~Teitelboim.
\newblock {Splitting of the Maxwell tensor - radiation reaction without
  advanced fields}.
\newblock {\em Phys. Rev.}, D1:1572--1582, 1970.
\newblock [Erratum: Phys. Rev.D2,1763(1970)].

\bibitem{Henneaux:2018cst}
M.~Henneaux and C.~Troessaert.
\newblock {BMS group at spatial infinity: the Hamiltonian (ADM) approach}.
\newblock {\em JHEP}, 03:147, 2018, 1801.03718.

\bibitem{Henneaux:2018gfi}
M.~Henneaux and C.~Troessaert.
\newblock {Asymptotic symmetries of electromagnetism at spatial infinity}.
\newblock 2018, 1803.10194.


\bibitem{Henneaux:2018hdj} 
  M.~Henneaux and C.~Troessaert,
 \newblock {Hamiltonian structure and asymptotic symmetries of the Einstein-Maxwell system at spatial infinity}.
  \newblock{arXiv:1805.11288 [gr-qc]}.
  
 \bibitem{Henneaux:2019yax} 
  M.~Henneaux and C.~Troessaert,
 \newblock{The asymptotic structure of gravity at spatial infinity in four spacetime 
dimensions}.
   \newblock{arXiv:1904.04495 [hep-th]}
   
\bibitem{BGP2} 
  C.~Bunster, A.~Gomberoff and A.~P\'erez,
  ``Hamiltonian analysis of the electromagnetic field on asymptotically null spacelike
   surfaces in the presence of a cosmological constant,'' \textit{in preparation}.
   
  \bibitem{Bondi:1960jsa}
H.~Bondi.
\newblock {Gravitational waves in General Relativity}.
\newblock {\em Nature}, 186(4724):535--535, 1960.


\bibitem{Regge:1957td}
T.~Regge and J.~A.~Wheeler.
\newblock {Stability of a Schwarzschild singularity}.
\newblock {\em Phys. Rev.}, 108:1063--1069, 1957.

\bibitem{Henneaux:2004jw} 
M.~Henneaux and C.~Teitelboim.
\newblock {Duality in linearized gravity}.
\newblock {\em Phys. Rev.}, D71:024018, 2005, gr-qc/0408101.

\bibitem{Bunster:2006rt}
C.~Bunster, S.~Cnockaert, M.~Henneaux, and R.~Portugues.
\newblock {Monopoles for gravitation and for higher spin fields}.
\newblock {\em Phys. Rev.}, D73:105014, 2006, hep-th/0601222.

\bibitem{Teitelboim:1973yj}
C.~Teitelboim.
\newblock {The Hamiltonian structure of space-time}.
\newblock {\em Ph.D. Thesis, Princeton, unpublished.}, 1973.

\bibitem{Teitelboim:1972vw}
C.~Teitelboim.
\newblock {How commutators of constraints reflect the space-time structure}.
\newblock {\em Annals Phys.}, 79:542--557, 1973.


\end{thebibliography}
\end{document}